\documentstyle[12pt]{article}
\newcommand{\beq}{\begin{equation}}
\newcommand{\eeq}{\end{equation}}
\newcommand{\ba}{\begin{eqnarray}}
\newcommand{\ea}{\end{eqnarray}}

\newcommand{\dsl}
  {\kern.06em\hbox{\raise.15ex\hbox{$/$}\kern-.56em\hbox{$\partial$}}}

\newcommand{\eeqarr}{\end{eqnarray}}
\newcommand{\ZZ}{{\rm \kern 0.275em Z \kern -0.92em Z}\;}
\begin{document}
\begin{titlepage}
\begin{center}
{\Huge Chern-Simons Theories}
\\
\vspace*{1cm}
{\Huge in the AdS/CFT Correspondence}
\\
\vspace*{1cm}
{\large Pablo Minces\footnote{pablo@fma.if.usp.br} and 
Victor O. Rivelles\footnote{rivelles@fma.if.usp.br}}
\\
\vspace*{0.5cm}
Universidade de S\~ao Paulo, Instituto de F\'{\i}sica\\
Caixa Postal 66.318 - CEP 05315-970 - S\~ao Paulo - Brazil
\vspace*{1cm}
\end{center}

\begin{abstract}
We consider the AdS/CFT correspondence for theories with a
Chern-Simons term in three dimensions. We find the two-point functions of
the boundary conformal field theories for the Proca-Chern-Simons theory
and
the Self-Dual model. We also discuss particular limits
where we find
the two-point function of the boundary conformal field theory for
the Maxwell-Chern-Simons theory.  In particular our results
are consistent with the equivalence between the Maxwell-Chern-Simons
theory and the Self-Dual model. 
\end{abstract}

\vskip 1cm

\begin{flushleft}
PACS numbers: 11.10.Kk 11.25.Mf\\
Keywords: AdS/CFT Correspondence, Chern-Simons Theory
\end{flushleft}
\end{titlepage}

\section{Introduction}

Since the proposal of Maldacena's conjecture 
that the large N limit of a certain conformal field theory (CFT) 
in a d-dimensional space is a boundary theory of string/M--theory
on $AdS_{d+1}\times K$ (where $K$ is a suitable compact space)
\cite{Malda}, an intensive work has been devoted to understand all
of its implications. 
In particular, a precise form to the conjecture has been given in
\cite{Polyakov}\cite{Witten}. Their suggestion is that the partition 
function for a field theory on $AdS_{d+1}$, considered as the
functional of the 
asymptotic value of the field on the boundary, is the generating 
functional for the correlation functions in the CFT on the boundary.
Schematically,
\beq
Z _{AdS}[\phi _{0}] = \int _{\phi _{0}}{\cal D}\phi\
\exp\left(-I[\phi]\right)\equiv Z _{CFT}[\phi _{0}] = \left<\exp\left(\int
_{\partial\Omega}d^{d}x{\cal O}\phi _{0}\right)\right>
\label{partition}
\eeq  
where $\phi _{0}$ is the boundary value of $\phi$ which couples to 
the boundary CFT operator ${\cal O}$. This allows us to obtain the
correlation functions of the boundary CFT theory in d dimensions by
calculating the
partition function on the $AdS _{d+1}$ side.

The AdS/CFT correspondence has been studied for
scalar fields \cite{Viswa2}, massive vector fields
\cite{Viswa}\cite{l'Yi}, spinor fields
\cite{Viswa}\cite{Sfetsos}\cite{Kaviani},
Rarita-Schwinger field \cite{Volovich}, classical gravity \cite{Liu} and
type IIB string theory \cite{Banks}\cite{Chalmers}. 
In this work we discuss the $AdS_{3}/CFT_{2}$ correspondence for vector
field theories including a Chern-Simons term. In section 2 we deal 
with the Proca-Chern-Simons theory. An explicit expression for the
two-point function of the boundary CFT is obtained. We then study the
massless limit to obtain the two-point function for the 
Maxwell-Chern-Simons theory \cite{tmgt}. In section 3 we deal with
the Self-Dual model \cite{self-dual}. Since the standard Self-Dual
action has at most first order derivatives it vanishes on-shell. Also
the variational principle requires a surface term in order to have a
stationary action \cite{henneaux}. As in the case involving spinors
\cite{Viswa}\cite{Sfetsos}\cite{Kaviani}\cite{Volovich} the only
contribution to the boundary CFT comes from a surface term. 
An expression for the two-point function on the border 
is then obtained. There is a well known equivalence
between the Self-Dual model and  the Maxwell-Chern-Simons
theory \cite{Deser}. We find that the resulting two-point functions
of the corresponding boundary CFT's are consistent with this
equivalence. Finally section 4 presents our conclusions. 

\section{The Proca-Chern-Simons Theory}

Since we are going to consider the Euclidean version of $AdS _{3}$ we
start with the Euclidean signature action for the Proca-Chern-Simons
theory which is given by 
\beq
I_{PCS} = - \int d^3x \, \sqrt{g} \left( \frac{1}{8}F
  _{\mu\nu}F^{\mu\nu} + \frac{1}{4}m^{2}A _{\mu}A^{\mu} + 
\frac{1}{\sqrt{g}}\frac{i\mu}{8}\epsilon^{\mu\nu\alpha}F
_{\mu\nu}A_{\alpha}\;+\;c.c. \right),
\label{1}
\eeq 
where $F _{\mu\nu} = \partial _{\mu}A _{\nu}-\partial _{\nu}A
_{\mu}$ and $\epsilon^{\mu\nu\alpha}$ is the Levi--Civita tensor
density with $\epsilon^{012}=1$. Using the variational principle to
obtain the field equations a surface term is generated
\beq
\int d^3x \; \partial_\mu \left( -\frac{1}{2} \sqrt{g} F^{\mu\nu}
  \delta A_\nu + i \frac{\mu}{8} \epsilon^{\mu\nu\alpha} A_\nu \delta
  A_\alpha  \;+\;  c.c.  \right).
\label{surface1}
\eeq
We will choose coordinates such that the Minkowski border of $AdS_3$
is situated at $x_0 = 0$. Then the  
boundary term Eq.(\ref{surface1}) will depend only on variations of the
spatial components $A_i$ of the vector potential. Choosing  boundary
conditions only on the $A_i$'s makes the boundary term to vanish so that no
further surface terms need to be added to the action
Eq.(\ref{1}). This kind of consideration will play a fundamental role
in the next section. 

The field equations which follow from Eq.(\ref{1}) are
\beq
\nabla_{\mu} F^{\mu\nu} - m^2 A^{\nu} - i \mu \frac{1}{\sqrt{g}} 
\epsilon^{\nu\alpha\beta}\partial_{\alpha}A_{\beta} = 0,
\label{11}
\eeq
which implies 
\beq
\nabla_{\mu}A^{\mu}=0.
\label{div}
\eeq
Solving Eqs.(\ref{11},\ref{div}) in the $AdS _{3}$ background is difficult
due to the presence of the Levi--Civita tensor density. However it can
be eliminated in the following way. Using Eq.(\ref{div}) in
Eq.(\ref{11}) we get
\beq
\left( \nabla^2 - m^2 - \frac{R}{3} \right) A^\mu - i \mu {}^*F^\mu =
0, 
\label{12}
\eeq
where ${}^*F^\mu = \frac{1}{2}\frac{1}{\sqrt{g}} \epsilon^{\mu\nu\alpha}
F_{\nu\alpha}$ and $R$ is the scalar curvature of $AdS _{3}$. Now
multiplying Eq.(\ref{11}) by the Levi--Civita 
tensor density and using again Eq.(\ref{div}) we get
\beq
\left( \nabla^2 - m^2 - \mu^2 - \frac{R}{3} \right) {}^* F^\mu + i \mu
m^2 A^\mu = 0.
\label{13}
\eeq
Finally eliminating ${}^* F^\mu$ from Eqs.(\ref{12},\ref{13}) we
arrive at 
\beq
\left( \nabla^2 - m_+^2 - \frac{R}{3} \right) \left( \nabla^2 - m_-^2
  - \frac{R}{3} \right) A^\mu = 0,
\label{14}
\eeq
where 
\beq
m^2_\pm(m,\mu) = \left[ \left( m^2 + \frac{\mu^2}{4} \right)^\frac{1}{2}
  \pm \frac{\mu}{2} \right]^2.
\label{masses}
\eeq
In the flat space limit this is an
indication that the Proca-Chern-Simons theory describes two excitations
with masses $m_\pm$ \cite{Paul}. We notice that the
solutions of Eq.(\ref{14}) must satisfy 
\beq
\left( \nabla^2 - m_+^2 - \frac{R}{3} \right) A^\mu = 0, 
\label{15}
\eeq
or
\beq
\left( \nabla^2 - m_-^2 - \frac{R}{3} \right) A^\mu = 0.
\label{16}
\eeq
Therefore the general solution of Eq.(\ref{14}) is a superposition of
solutions of the Proca theory with masses $m_+$ and $m_-$. The Proca
theory has been analyzed in \cite{Viswa} and we now follow closely the
derivation in that paper.  

We take the usual representation of the $AdS _{3}$ described by the
half space $x _{0}>0$, $x _{i} \in {\bf R}$ with metric
\beq
ds^{2}=\frac{1}{x _{0}^{2}} \sum_{\mu=0}^2 dx^{\mu}dx^{\mu}, 
\label{2} 
\eeq
for which the curvature scalar is $R=-6$. 
As in \cite{Viswa} we also introduce vector potentials with Lorentz 
indices $\tilde A_{\mu}$ using the vielbein of $AdS _{3}$ 
\beq
{\tilde A_{\mu}} = x_{0} A_\mu. 
\label{4}
\eeq
The solutions which are regular at $x_0 \rightarrow \infty$ can be written as 
\beq
{\tilde A _{\mu}} = \frac{1}{2}\left( {\tilde A _{\mu}^{+}} + {\tilde
A _{\mu}^{-}}\right), 
\label{112}
\eeq
where
\beq
{\tilde A_{0}^{\pm}}(x) = \int\frac{d^{2}k}{(2\pi)^{2}} e^{-i
\vec{k}\cdot\vec{x}} \; x _{0}^{2} \; a _{0}^{\pm}(\vec{k}) \; K
_{m_{\pm}}(k x_0),
\label{113}
\eeq
\beq
{\tilde A_{i}^{\pm}}(x)  = \int\frac{d^{2}k}{(2\pi)^{2}}e^{-i
\vec{k}\cdot\vec{x}} \: x _{0} \; \left( a_{i}^{\pm}(\vec{k}) \; 
K_{m_{\pm}}(k x_0) + i a_{0}^{\pm}(\vec{k}) \; \frac{k _{i}}{k} \; 
x_{0} \; K_{m_{\pm} + 1}(k x_0) \right),
\label{114}
\eeq
$\vec{x}=(x^{1},x^{2})$, $k = \mid\vec{k}\mid$,
$K_{m_\pm}$ are the modified Bessel functions, and
from now on
$m_\pm$ is to be understood as $\mid m_\pm\mid$. The normalization in 
Eq.(\ref{112}) has been chosen so that it reproduces the results
in \cite{Viswa} in the particular case $\mu=0$ (and hence ${\tilde A^{+}}
= {\tilde A^{-}}$). Inserting Eqs.(\ref{113},\ref{114}) in the
original equations of motion Eq.(\ref{11}) gives the following relations
among the coefficients $a^{\pm}$
\beq
\mu m_{\pm} a_{i}^{\pm}(\vec{k}) = \mp i \mu
m_{\pm}\epsilon^{0ij}a_{j}^{\pm}(\vec{k}),
\label{rel1}
\eeq
\beq
\mu m_{\pm} a_{0}^{\pm}(\vec{k}) \left( \mp\epsilon^{0ij}k_{j} - i
k_{i}\right) = 
\mp i \mu k^{2} \epsilon^{0ij}a_{j}^{\pm}(\vec{k}).  
\label{rel2}
\eeq
From Eq.(\ref{div}) we also find 
\beq
i k_i a^\pm_i(\vec{k}) = m_\pm a_0^\pm(\vec{k}),
\label{111c}
\eeq
which is consistent with Eq.(\ref{rel2}). We consider first the case
$\mu\not= 0$,$m\not= 0$ and rewrite Eqs.(\ref{rel1}) and
(\ref{rel2})
as 
\beq
a_{i}^{\pm}(\vec{k}) = \mp i\epsilon^{0ij}a_{j}^{\pm}(\vec{k}),
\label{rel3}
\eeq
\beq
m_{\pm} a_{0}^{\pm}(\vec{k}) \left( \pm i k_{i} +
\epsilon^{0ij}k_{j} \right)  = \mp \; k^{2} \; a_{i}^{\pm}(\vec{k}).
\label{rel4}
\eeq
In order to capture the effect of the Minkowski boundary of the $AdS_3$,
situated at $x_0=0$, 
we first consider a Dirichlet boundary value problem on the boundary
surface $x_0=\epsilon > 0$ and 
then take the limit $\epsilon \rightarrow 0$. The potential at the
near boundary surface will be denoted by $\tilde{A}_{\epsilon,
  \mu}$. Imposing the near boundary condition on
Eqs.(\ref{113},\ref{114}) and using Eqs.(\ref{rel3},\ref{rel4}) allow us
to find the coefficients  $a^\pm$ in terms of the Fourier transform of 
the fields $\tilde{A}_{\epsilon, i}$  
\beq
a _{0}^{\pm}(\vec{k})  =
\pm\frac{\epsilon^{0ij}\;\omega_{i}^{\mp}(\vec{k})\;{\tilde
A}_{\epsilon,j}(\vec{k})}{\epsilon^{0ij}\;\omega_{i}^{-}(\vec{k})\;
\omega_{j}^{+}(\vec{k})},
\label{114a}
\eeq
\ba
a_i^\pm (\vec{k})& = &  \frac{\epsilon^{-1}\left( {\tilde A}_{\epsilon,i}
    (\vec{k})\mp i \epsilon^{0ij} {\tilde A}_{\epsilon,j}
    (\vec{k})\right)}{K_{m_{\pm}}(k\epsilon)} + \nonumber \\ 
 &+& \frac{k_i \mp i \epsilon^{0ij} k_{j}}{k} \; 
\frac{i\epsilon}{2K_{m_{\pm}}(k\epsilon)}\;\frac{\epsilon^{0kl}{\tilde
A}_{\epsilon,l}(\vec{k})}{\epsilon^{0rs}\;\omega_{r}^{-}(\vec{k})\;
\omega_{s}^{+}(\vec{k})} \times \nonumber \\
& \times & \left[\omega_{k}^{+}(\vec{k})K_{m_{-}+1}(k\epsilon)-
\omega_{k}^{-}(\vec{k})K_{m_{+}+1}(k\epsilon)\right],
\label{114b}
\ea
where
\beq
\omega_{i}^{\pm}(\vec{k}) = \frac{i\epsilon}{2k^{2}}\;\left[
m_{\pm}\left(
k_{i}\pm
i\epsilon^{0ij}k_{j}\right)K_{m_{\pm}}(k\epsilon)+k_{i}k
\epsilon K_{m_{\pm}-1}(k\epsilon)\right].
\label{114c}
\eeq
From Eqs.(\ref{113}) and (\ref{114}) we get
\beq
{\tilde F}_{\epsilon,0i}^{\pm}(\vec{x}) = \left(
1-m_{\pm}\right)\;\frac{1}{\epsilon}\;{\tilde
A}_{\epsilon,i}^{\pm}(\vec{x}) 
\;-\;\int\frac{d^{2}k}{\left( 2\pi\right)^{2}}e^{-i\vec{k}\cdot\vec{x}}
a_{i}^{\pm}(\vec{k})k\epsilon K_{m_{\pm}-1}(k\epsilon).
\label{efe}
\eeq
Using this we can finally calculate 
the value of the classical action in the near boundary surface using the
action Eq.(\ref{1}). After an 
integration by parts and using the equations of motion we find that
there is only a contribution from the boundary 
\beq
I_{PCS} = - \frac{1}{4} \int d^3x \; \partial_\mu ( \sqrt{g}
F^{\mu\nu} A_\nu )\;+\;c.c., 
\label{boundary}
\eeq
which evaluated on the near boundary surface gives
\beq
I _{PCS} = -\frac{1}{4}\int d^2x \; \epsilon^{-2}{\tilde A
_{\epsilon,i}}\left(
-{\tilde A _{\epsilon,i}} + \epsilon{\tilde F
_{\epsilon,0i}}\right)\;+\;c.c.
\label{115}
\eeq 
Using Eqs.(\ref{rel3},\ref{114b},\ref{efe},\ref{115}), keeping only
the relevant terms in the series expansion of the Bessel functions, 
and integrating over the momenta  we get
\beq
I _{PCS} = I^{+} + I^{-} + \cdots,
\label{116}
\eeq
where the dots stand for either contact terms 
or higher order terms in $\epsilon$ and 
\ba
I^{\pm} \!\!&=&\!\! \frac{m_{\pm}(m,\mu)}{8}\int d^2x\;
\epsilon^{-2}\left[{\tilde A
_{\epsilon,i}}(\vec{x}) \; {\tilde A
_{\epsilon,i}^{\pm}}(\vec{x})\; + \;c.c.\;\right]
\nonumber\\&-&
\frac{1}{4}\;{\tilde c _{\pm}(m,\mu)}{\tilde \Delta _{\pm}}(m,\mu)
\int d^2x d^2y \; \left[{\tilde A
_{\epsilon,i}}(\vec{x}) \;
{\tilde A _{\epsilon,i}}(\vec{y})\; + \;c.c.\;\right]
\;\frac{\epsilon^{2\left[m_{\pm}(m,\mu) - 1\right]}}
{\mid\vec{x} - \vec{y}\mid^{2{\tilde \Delta _{\pm}}(m,\mu)}}\nonumber\\
&+&\;{\tilde c _{\pm}(m,\mu)}{\tilde \Delta _{\pm}}(m,\mu)
\int d^2x d^2y \;
\left[ {\tilde A
_{\epsilon,i}^{R}}(\vec{x}){\tilde A   
_{\epsilon,j}^{R}}(\vec{y})\;-\;{\tilde A   
_{\epsilon,i}^{I}}(\vec{x}){\tilde A   
_{\epsilon,j}^{I}}(\vec{y})
\right.\nonumber\\
&&
\left.
\mp\;\epsilon^{0il}\;\left(
{\tilde A   
_{\epsilon,l}^{R}}(\vec{x}){\tilde A   
_{\epsilon,j}^{I}}(\vec{y})\;+\;{\tilde A   
_{\epsilon,l}^{I}}(\vec{x}){\tilde A   
_{\epsilon,j}^{R}}(\vec{y})\right)\right]
\;\frac{\epsilon^{2
\left[m_{\pm}(m,\mu) - 1\right]}}
{\mid\vec{x} - \vec{y}\mid^{2{\tilde \Delta _{\pm}}(m,\mu)}}\;
\frac{(x-y)_{i}(x-y)_{j}}{\mid\vec{x}-\vec{y}\mid^{2}}\;,\nonumber\\
\label{117}
\ea
\beq
{\tilde \Delta _{\pm}}(m,\mu) = m _{\pm}(m,\mu) + 1,
\label{131}
\eeq
\beq
{\tilde c _{\pm}}(m,\mu) = \frac{m _{\pm}(m,\mu)}{\pi}\;.
\label{132}
\eeq
Here ${\tilde A}_{\epsilon,i}^{R} \;\left( {\tilde
A}_{\epsilon,i}^{I}\right)$ denotes the real (imaginary) part of
${\tilde A}_{\epsilon,i}$.
Since the metric is singular in the border the
action is divergent and 
the limit $\epsilon\rightarrow 0$ has to be taken carefully
\cite{Freedman}. In order to have a finite action we take the limit
\beq
\lim _{\epsilon\rightarrow 0}\epsilon^{m_{-}(m,\;\mid\mu\mid)
- 1}{\tilde
A_{\epsilon,i}}(\vec{x}) = {\tilde A_{0,i}}(\vec{x}).
\label{133}
\eeq
Then we use the equivalence AdS/CFT in the form
\beq
\exp\left( -I _{AdS}\right) \equiv \left<\exp\left(\int d^2x \;  
J_{i}(\vec{x}) \; A _{0,i}(\vec{x})\right)\right>.
\label{134}
\eeq
When $\mu < 0$ we have $m_{-}(m,\;\mid\mu\mid) - 1 =
m_{+}(m,\mu) - 1$, the relevant part of $I^{-}$ vanishes and
the only contribution to the two-point function of the conformal field
$J_{i}^{PCS}$ on the boundary CFT comes from $I^{+}$. When $\mu > 0$
we have $m_{-}(m,\;\mid\mu\mid) - 1 = m_{-}(m,\mu) - 1$, and the only
contribution to the two-point function
of $J_{i}^{PCS}$ comes from $I^{-}$. In both cases we find the following
two-point function
\beq
\left <J _{i}^{PCS}(\vec{x}) \; J_{j}^{PCS}(\vec{y})\right> =
{\tilde c _{PCS}}{\tilde \Delta
_{PCS}}\left(
\delta_{ij}-2\frac{(x-y)_{i}(x-y)_{j}}{\mid\vec{x}-\vec{y}\mid^{2}}\right)
\mid\vec{x} - \vec{y}\mid^{-2{\tilde 
\Delta_{PCS}}},
\label{139}
\eeq
where
\beq
{\tilde \Delta_{PCS}} = {\tilde \Delta}_{-}(m,\;\mid\mu\mid),
\label{139'}
\eeq
and
\beq
{\tilde c _{PCS}} = {\tilde c _{-}}(m,\;\mid\mu\mid),
\label{139''}
\eeq
so that $J_{i}^{PCS}$ has conformal dimension ${\tilde \Delta_{PCS}}$. 
It is important to note that the identification Eq.(\ref{133}) agrees 
with the requirement 
that the isometries of $AdS_{3}$ correspond to the conformal isometries in
$CFT_{2}$ \cite{Freedman}.

Now we consider the particular cases $m = 0$ and $\mu = 0$. 
In order to get the boundary CFT associated to the Maxwell-Chern-Simons
theory we take $m = 0$ in Eq.(\ref{masses}), which gives
\beq
m_{\pm}(0,\mu) = \frac{1}{2}\left( \mid\mu\mid\pm\;\mu\right).
\label{masses2}
\eeq
When $\mu>0$ Eq.(\ref{rel2}) implies 
$a_{1}^{-}(\vec{k})=a_{2}^{-}(\vec{k})=0$ and the only contribution
to the action comes from $I^{+}$, whereas when $\mu<0$ Eq.(\ref{rel2})
fixes $a_{1}^{+}(\vec{k})=a_{2}^{+}(\vec{k})=0$ and the only contribution
comes from $I^{-}$. So the actions corresponding to the cases $\mu>0$
and $\mu<0$ read
\ba
I^{|\mu| = \pm \mu}_{MCS} \!\!&=&\!\! \frac{\mid\mu\mid}{8}\int d^2x\;
\epsilon^{-2}\left[{\tilde A
_{\epsilon,i}}(\vec{x}) \; {\tilde A
_{\epsilon,i}^{\pm}}(\vec{x})\; + \;c.c.\;\right]
\nonumber\\&-&
\frac{1}{4}\;{\tilde c _{MCS}}{\tilde \Delta _{MCS}}
\int d^2x d^2y \; \left[{\tilde A   
_{\epsilon,i}}(\vec{x}) \; 
{\tilde A _{\epsilon,i}}(\vec{y})\; + \;c.c.\;\right]
\;\frac{\epsilon^{2\left[\mid\mu\mid - 1\right]}}
{\mid\vec{x} - \vec{y}\mid^{2{\tilde \Delta _{MCS}}}}\nonumber\\
&+&\;{\tilde c _{MCS}}{\tilde \Delta _{MCS}}
\int d^2x d^2y \;
\left[ {\tilde A
_{\epsilon,i}^{R}}(\vec{x}){\tilde A
_{\epsilon,j}^{R}}(\vec{y})\;-\;{\tilde A
_{\epsilon,i}^{I}}(\vec{x}){\tilde A
_{\epsilon,j}^{I}}(\vec{y})
\right.\nonumber\\
&&
\left.
\mp\;\epsilon^{0il}\;\left(
{\tilde A
_{\epsilon,l}^{R}}(\vec{x}){\tilde A
_{\epsilon,j}^{I}}(\vec{y})\;+\;{\tilde A
_{\epsilon,l}^{I}}(\vec{x}){\tilde A
_{\epsilon,j}^{R}}(\vec{y})\right)\right]
\;\frac{\epsilon^{2
\left[\mid\mu\mid - 1\right]}}
{\mid\vec{x} - \vec{y}\mid^{2{\tilde \Delta _{MCS}}}}\;
\frac{(x-y)_{i}(x-y)_{j}}{\mid\vec{x}-\vec{y}\mid^{2}}\nonumber\\
&+& \cdots,
\label{117'}
\ea
where
\beq
{\tilde \Delta _{MCS}} = \mid\mu\mid + 1,
\label{140'}
\eeq
and
\beq
{\tilde c _{MCS}} = \frac{\mid\mu\mid}{\pi}.
\label{140''}
\eeq
We take 
\beq
\lim _{\epsilon\rightarrow 0}\epsilon^{\mid\mu\mid
- 1}{\tilde
A_{\epsilon,i}}(\vec{x}) = {\tilde A_{0,i}}(\vec{x}),
\label{identif}
\eeq
and use again the AdS/CFT correspondence Eq.(\ref{134}), so that in
both cases, $\mu>0$ and $\mu<0$, we get the following two-point
function for the boundary conformal field $J _{i}^{MCS}$
\beq
\left <J _{i}^{MCS}(\vec{x}) \; J _{j}^{MCS}(\vec{y})\right> =
{\tilde c _{MCS}}{\tilde \Delta
_{MCS}}\left( \delta
_{ij} - 2\frac{(x-y) _{i}(x-y) _{j}}{\mid\vec{x} -
\vec{y}\mid^{2}}\right)\mid\vec{x} - \vec{y}\mid^{-2{\tilde \Delta  
_{MCS}}},
\label{140}
\eeq
so that $J _{i}^{MCS}$ has conformal dimension ${\tilde
\Delta _{MCS}}$. As it is well known the
Maxwell-Chern-Simons theory describes a particle with 
mass $\mu$ \cite{tmgt} and
this fact is reflected in the conformal dimension Eq.(\ref{140'}).
Furthermore, our result is consistent with the holographic principle
since the mass $m_{-}(0,\mid\mu\mid) = 0$ is not physical in the bulk
\cite{tmgt} and does not contribute to the border two-point function. 

The Proca theory has been considered in \cite{Viswa} and we derive 
here the main results for completeness. Making $\mu = 0$ on
Eq.(\ref{masses}) gives $m_{\pm}(m,0) = m$ so that $A_{\mu}^{+} = 
A_{\mu}^{-} = A_{\mu}$. Eqs.(\ref{rel1},\ref{rel2}) vanish 
identically and the field $ A_{\mu}$ is real. The action reads
\ba
I_{P} &=& \frac{m}{2}\int d^{2}x\;\epsilon^{-2}
  {\tilde A}_{\epsilon,i}(\vec{x}) {\tilde A}_{\epsilon,i}(\vec{x})
\nonumber\\
  &-&\; {\tilde c}_{P}{\tilde \Delta}_{P}
  \int d^2x d^2y {\tilde
A}_{\epsilon,i}(\vec{x}){\tilde A}_{\epsilon,i}(\vec{y})
  \frac{\epsilon^{2(m-1)}}{|\vec{x}-\vec{y}|^{2{\tilde \Delta}_{P}}}    
  \left(\delta_{ij} - 2
  \frac{(x-y)_i(x-y)_j}{|\vec{x}-\vec{y}|^2} \right)\nonumber\\&+&\cdots,
\label{ac}
\ea
where
\beq 
{\tilde \Delta _{P}} = m + 1,
\label{141'}
\eeq
and
\beq
{\tilde c _{P}} = \frac{m}{\pi}.
\label{141''}
\eeq
Taking
\beq
\lim _{\epsilon\rightarrow 0}\epsilon^{m - 1}{\tilde
A_{\epsilon,i}}(\vec{x}) = {\tilde
A_{0,i}}(\vec{x}),
\label{identif2}
\eeq
and using the AdS/CFT correspondence Eq.(\ref{134}) we get
\beq
\left <J _{i}^{P}(\vec{x}) \; J _{j}^{P}(\vec{y})\right> =
2{\tilde c _{P}}{\tilde \Delta
_{P}}\left( \delta
_{ij} - 2\frac{(x-y) _{i}(x-y) _{j}}{\mid\vec{x} -
\vec{y}\mid^{2}}\right)\mid\vec{x} - \vec{y}\mid^{-2{\tilde \Delta
_{P}}},
\label{141}  
\eeq
so that the field $J _{i}^{P}$  has conformal dimension ${\tilde \Delta
_{P}}$.
\section{The Self-Dual Model}

We now start with the Euclidean signature action
\beq
I^0_{SD}= - \int d^3x \; \sqrt{g} \left( \frac{1} {\sqrt{g}}
  \frac{i\kappa}{8 } \epsilon^{\mu\nu\alpha} F_{\mu\nu}A_{\alpha} +
  \frac{1}{4} M^{2} A_{\mu} A^{\mu}\;+\;c.c. \right),
\label{145}
\eeq
for the Self-Dual model \cite{self-dual}. In order to have
a stationary action we must supplement the action Eq.(\ref{145}) 
with a surface term which cancels its variation \cite{henneaux}. The
variational principle generates a boundary term 
\beq
-\frac{\kappa}{2}\int d^{2}x\;\epsilon^{0ij}
\;\left[ A_{i}^{R}(\vec{x})\delta
A_{j}^{I}(\vec{x})\;+\;A_{i}^{I}(\vec{x})\delta
A_{j}^{R}(\vec{x})\right], 
\label{ac2}
\eeq
which is written in terms of the real and imaginary parts of the vector
potential. Since the field equations derived from Eq.(\ref{145}) are
first order differential equations we can not choose boundary
conditions which fix simultaneously the real and imaginary parts of
the $A_i$'s. Then we choose boundary conditions on the $A_i^R$'s
leaving a non-vanishing term proportional to the $\delta A_i^I$'s in
the boundary term Eq.(\ref{ac2}). So we add to the action
Eq.(\ref{145}) a surface term of the form 
\beq
I^{surface}_{SD} = \frac{\kappa}{2}\int d^{2}x\;\epsilon^{0ij}
\;A_{i}^{R}(\vec{x})A_{j}^{I}(\vec{x}),
\label{ac5}
\eeq
and the action 
\beq
I_{SD} = I^{0}_{SD} + I^{surface}_{SD},
\label{ac4}
\eeq
is now stationary. 

The field equations which follow from the action Eq.(\ref{ac4}) are 
\beq
i \kappa \frac{1}{\sqrt{g}} \epsilon^{\nu\alpha\beta}
\partial_{\alpha}A_{\beta} + M^2 A^{\nu} = 0.
\label{1000}
\eeq
It implies again 
\beq
\nabla_{\mu}A^{\mu}=0.
\label{1001}
\eeq
Using the equations of motion we find that 
\beq
A_{j}^{I}=-\frac{\kappa}{2M^{2}}\;\sqrt{g}\;\epsilon^{j\alpha\beta}
F^{R,\;\alpha\beta},
\label{ac6}
\eeq
so that $I^{surface}_{SD}$ is rewritten as
\ba
I^{surface}_{SD} &=& - \frac{\kappa^{2}}{2M^{2}} 
\int d^3x \; \partial_\mu (\sqrt{g}
F^{R,\;\mu\nu} A_{\nu}^{R} )\nonumber\\
&=& -\frac{\kappa^{2}}{2M^{2}} \int
d^2x  \;\epsilon^{-2}{\tilde A_{\epsilon,i}^{R}} \left(-{\tilde
    A_{\epsilon,i}^{R}} + \epsilon{\tilde F_{\epsilon,0i}^{R}}
\right), 
\label{ac7}
\ea
and depends only on the $\tilde{A}^R_{\epsilon, i}$'s. 

As in the case of the Proca-Chern-Simons theory we can eliminate the
Levi--Civita tensor density by increasing the order of the equations of
motion. We then get
\beq
\left( \nabla^2 - \frac{M^4}{\kappa^{2}} - \frac{R}{3} \right)
A^\mu = 0.
\label{1002}
\eeq
Proceeding as before we find the solution
\beq
{\tilde A _{0}}(x) = \int \frac{d^{2}k}{(2\pi)^{2}} \; 
e^{-i \vec{k} \cdot \vec{x}} \; x _{0}^{2} \; b _{0}(\vec{k}) \;
K_{\frac{M^{2}}{\mid\kappa\mid}}(k x_0), 
\label{145'}
\eeq
and
\beq
{\tilde A_{i}}(x) = \int\frac{d^{2}k}{(2\pi)^{2}} \; 
e^{-i\vec{k} \cdot \vec{x}} \; x_{0} \; \left( b _{i}(\vec{k}) \;
  K_{\frac{M^{2}}{\mid\kappa\mid}}(k x_0) + ib _{0}(\vec{k})
\frac{k_{i}}{k}x _{0} \;
  K_{\frac{M^{2}}{\mid\kappa\mid} +1}(k x_0) \right).
\label{145''}
\eeq
From Eq.(\ref{1001}) we get
\beq
i k_i b_i(\vec{k}) = \frac{M^{2}}{\mid\kappa\mid} b_0(\vec{k}). 
\label{153''}
\eeq

As before we would like to express the coefficients $b$ in terms of the
Fourier components of the
vector field at the near boundary surface $x_0 = \epsilon$. It should
be noted that since the bulk term of the action Eq.(\ref{ac4}) vanishes
on-shell all the contributions to the two-point function on the boundary
CFT come from the surface term Eq.(\ref{ac7}) and  depend 
only on the real components $\tilde{A}_{\epsilon,i}^R$'s. Using that
Eq.(\ref{1002}) applies separately to the real and imaginary parts of
$A_\mu$ we find for the relevant parts of the coefficients $b$ (i.e.,
those which contain the real components of the $\tilde{A}_{\epsilon,
  i}$'s) the following expressions 
\beq
b_{0}(\vec{k}) = \frac{-i\epsilon^{-1}{\tilde A
_{\epsilon,i}^{R}}(\vec{k}) k_{i}}{\frac{M^2}{\mid\kappa\mid}
K_{\frac{M^{2}}{\mid\kappa\mid}}(k \epsilon) + k\epsilon
K_{\frac{M^{2}}{\mid\kappa\mid}
-1}(k \epsilon)}, 
\label{154}
\eeq
\beq
b_{i}(\vec{k})  = \frac{\epsilon^{-1} {\tilde
    A_{\epsilon,i}^{R}}(\vec{k})}{K_{\frac{M^{2}}{\mid\kappa\mid}}(k
\epsilon)} - 
\frac{k _{i}k_{j}}{k} \frac{{\tilde A _{\epsilon,j}^{R}}(\vec{k}) 
K_{\frac{M^{2}}{\mid\kappa\mid} +1}(k
\epsilon)}{\frac{M^2}{\mid\kappa\mid}
K_{\frac{M^{2}}{\mid\kappa\mid}}^{2}(k \epsilon) + k \epsilon
K_{\frac{M^{2}}{\mid\kappa\mid}}(k \epsilon) 
K_{\frac{M^{2}}{\mid\kappa\mid} - 1}(k \epsilon)}. 
\label{155}
\eeq
Proceeding as before we find
\ba
I _{SD}\!\!&=&\!\!\frac{\mid\kappa\mid}{2}
\int d^2x
\;
\epsilon^{-2} \; {\tilde A_{\epsilon,i}^{R}}(\vec{x}) \; {\tilde A
  _{\epsilon,i}^{R}}(\vec{x}) 
\nonumber\\&-& \;{\tilde c}_{SD}{\tilde
\Delta}_{SD}
\int d^2x d^2y \; {\tilde A_{\epsilon,i}^{R}}(\vec{x}) \; 
{\tilde A_{\epsilon,j}^{R}}(\vec{y}) \; \frac{\epsilon^{2\left(
\frac{M^{2}}{\mid\kappa\mid}-1\right)}}
{\mid\vec{x} - \vec{y}\mid^{2{\tilde \Delta}_{SD}}}\left( \delta
_{ij} - 2\frac{(x-y) _{i}(x-y) _{j}}{\mid\vec{x} -
\vec{y}\mid^{2}}\right)\nonumber\\&+& ... , 
\label{157}
\ea
where
\beq
{\tilde \Delta}_{SD} = \frac{M^{2}}{\mid\kappa\mid} + 1, 
\label{158}
\eeq
and
\beq
{\tilde c}_{SD} = \frac{\mid\kappa\mid}{\pi}\;. 
\label{159}
\eeq
Now taking 
\beq
\lim_{\epsilon\to
0}\epsilon^{\frac{M^{2}}{\mid\kappa\mid} -
1}{\tilde A _{\epsilon,i}^{R}}(\vec{x}) = A _{0,i}(\vec{x}),
\label{identif3}
\eeq
and using
the AdS/CFT correspondence Eq.(\ref{134}) we find the two-point function
of the 
conformal field $J_{i}^{SD}$ coupled to the field ${\tilde A_{i}}$ on the
boundary 
\beq
\left <J_{i}^{SD}(\vec{x})\; J_{j}^{SD}(\vec{y})\right> = 2
{\tilde c}_{SD}{\tilde
\Delta}_{SD}\left( \delta
_{ij} - 2\frac{(x-y) _{i}(x-y) _{j}}{\mid\vec{x} -
\vec{y}\mid^{2}}\right)\mid\vec{x} - \vec{y}\mid^{-2{\tilde \Delta}_{SD}}. 
\label{162}
\eeq
We then find that the field $J_{i}^{SD}$ has conformal dimension ${\tilde
\Delta}_{SD}$. 

Comparing Eqs.(\ref{158}) and (\ref{140'}) we see that
the conformal dimensions of the conformal fields corresponding to the
Maxwell-Chern-Simons theory and the Self-Dual model are the same for
$\frac{M^{2}}{\mid\kappa\mid} = \mid\mu\mid$ in agreement with the
equivalence between those models \cite{Deser}.

\section{Conclusions}

As expected from the holographic principle the conformal dimensions
depend only on the masses of the corresponding theories in the
bulk. Although the solutions in the bulk expressed in terms of the
boundary values have a complicated form the boundary two--point
functions are very simple as dictated by conformal invariance. In fact
for each theory 
we could have just solved the corresponding Proca equations and used
this solution to find the two--point function on the border. The
final result is insensitive to the detailed structure in the bulk. 

Another manifestation of the holographic principle is the fact that in
the massless limit the Proca-Chern-Simons theory gives rise to only
one massive excitation of mass $|\mu|$ and the massless mode becomes
unphysical. This is reflected in the border CFT where the two-point
function Eq.(\ref{140}) has a contribution only from the massive mode
of the bulk theory. 

The equivalence between the Self-Dual model and the
Maxwell-Chern-Simons theory in flat space-time is well known
\cite{Deser} and it can easily be shown to be true also in curved
space-time either at the level of the equations of motion or by
defining a master Lagrangian in curved space-time. The fact that we
obtain the same conformal dimension for 
the corresponding CFT's in the border is in support of the
holographic principle. Not only the conformal dimensions are the
same but the coefficients $\tilde{c}$ of the two-point functions can
be made the same by an appropriate normalization of the Self-Dual
action. Since we started with two independent parameters in
Eq.(\ref{145}) we can now choose $M = | \kappa|$ so that the model
describes a particle with mass $M$. Now our results have an universal
form in which the conformal dimension and the two-point function
coefficient can be written as $\tilde{\Delta} = m +1$ and  $\tilde{c} =
m/\pi$ respectively, where $m$ is the mass of the bulk theory.  

\section{Acknowledgements}

It is a pleasure to thank Carlos N\'u\~nez for encouragement and very
useful conversations. P.M. acknowledges the support by CAPES. V.O.R. 
is partially supported by CNPq and acknowledges a grant by FAPESP.

\end{document}